\documentclass[aps,pra,showpacs,preprint,prl,preprint]{revtex4}
\def\beginwide{
        \end{multicols} \vspace*{-0.5cm} \noindent
        \rule{3.5in}{.1mm}\rule{.1mm}{5mm} \widetext \medskip } 

\def\endwide{
        \hspace*{3.5in}~\rule[-5mm]{.1mm}{5mm}\rule{3.5in}{.1mm}
        \begin{multicols}{2} \vspace*{-1.0cm} \noindent }   
\begin{document}
\title{Formation of Two Component Bose Condensate During The Chemical Potential Curve Crossing}
\author{M.A. Kayali, N.A. Sinitsyn}
\affiliation{Department of Physics, Texas A\&M University\\ College Station,
Texas 77843-4242, USA}
\date{\today}

\begin{abstract}
In this article we study the formation of the two modes Bose-Einstein condensate and the correlation between them. We show that beyond the mean field 
approximation the dissociation of a molecular condensate due to the
  chemical potential curve crossing leads to the formation of two modes condensate. We also show that these two resulting modes are correlated in a two mode 
squeezed state. 
\end{abstract}
\maketitle


\section{Introduction}
Recent studies on Bose-Einstein condensate have directed the attention of many theorists and experimentalists to the multimodes Bose-Einstein condensate. In particular the appearance of multimodes condensate from a molecular one via molecular condensate dissociation process. In \cite{first}, \cite{second} the authors studied the dissociation of a molecular condensate into a one mode atomic condensate. Moreover, Yurovsky {\it{et.al}} showed that the resulting atomic condensate is in a squeezed state. They also showed that the number of atoms and squeezing parameters in the case of time dependent chemical potentials depend on the rate of curve crossing and the coupling between the molecular and the atomic fields.\\
The problems of curve crossing can be divided into two different classes. These are the $r$-dependent crossings and the $t$-dependent ones. In the former the potentials are coordinates dependent and the crossing takes place in space, while in the latter the potentials are time dependent. The standard approach to the crossing problems is the semiclassical approach which is known as Landau-Zener (LZ) theory. The LZ theory provide a very good description of the crossing problem for systems with potentials that cross at one point only. However, if crossing between two potentials occur at two points or more then the solution of the problem is possible by using a different approach which is out of the LZ theory.\\ 

In the production of an atomic condensate from a molecuar one the curve crossing problem arise from the crossing of the chemical potentials of the condensates. In this article, we study the creation of multi modes atomic condensate from a molecular one near the curve crossing of the system's chemical potentials. We show that within reasonable approximations the problem can be 
reduced to the case of two atomic modes and one molecular mode condensates. We show that exact results for this problem are still possible since the equations for the operators evolution decouple so that the two level Landau-Zener problem can be exactly solved. Compared to one atomic mode process, the  two modes system provides much more possibilities for correlated states since the initial conditions depend on two complex parameters which describe the initial states of the two modes. The aim of this article is to generalize the results of \cite{first} to the case that includes exited modes of the atomic condensate.
In this article we study the process when the dissociation of atomic collisions in one mode leads to the production of atoms in two different modes.

\section{The Problem}

   In \cite{mm} it was shown that the description of molecular and atomic 
condensate by two modes only is generally not realistic, since the exited 
atomic modes interact with molecular condensate. Consequently some of them 
become highly populated and must be taken into account. Generally speaking the Hamiltonian that includes more than one atomic and molecular modes can be 
written as follows\cite{mm}:

\begin{eqnarray}
\hat{H} &=& \sum_{{\bf k}} (\mu_k a^{\dagger}_{{\bf
k}}a_{{\bf k}} +\nu_k\frac{1}{2}\psi^{\dagger}_{{\bf k}}\psi_{{\bf k}})
 +g_{k,k'} \sum_{{\bf k},{\bf k'}}
\psi_{{\bf k}+{\bf k'}}^{\dagger}a_{{\bf k}}a_{{\bf k'}} +\psi_{{\bf
k}+ {\bf k'}}a^{\dagger}_{{\bf k}}a^{\dagger}_{{\bf k'}} \nonumber\\
  &+& \frac{1}{2}g_{int}
\sum_{{\bf k},{\bf k'},{\bf k''}} a^{\dagger}_{{\bf k}+{\bf
k'}-{\bf k''}} a^{\dagger}_{{\bf k''}}a_{{\bf k'}}a_{{\bf k}}
\end{eqnarray} 

If we assume that the chemical potentials $\mu_{k}$ depend on time then the Landau-Zener model for multimode case becomes very complicated to solve even in the case when the molecular field is treated as a $c$-number. However, if the 
number of atoms in the atomic modes is not very large we can disregard the 
interaction term $(g_{int}=0)$. Let us consider a system that initially has one highly populated mode of the molecular condensate. For example, molecular mode with total momentum equals zero. Due to momentum conservation, the atomic mode 
with momentum $\vec{k}$ would couple via the interaction with the molecular mode to the atomic mode of momentum $-\vec{k}$ only. In such systems the problem is reduced to the interaction of two atomic modes with molecular condensate. Other molecular modes can also be considered but in this article we will show that the effective coupling constant in this case is propotional to the square root of the number of particles in the 
molecular mode; therefore, we can disregard the interactions with the initially empty molecular modes.

Let us consider a system of two atomic fields $A$ and $B$ and one molecular field AB. Suppose $\hat{a}$, $\hat{b}$ and $\hat{\psi}$ are
 the annihilation operators of $A$, $B$ and $AB$ field respectively. If we 
disregard the interaction between $A$ and $B$ during curve crossing then the 
Hamiltonian for molecular dissociation in the process $AB \rightarrow A+B$
   is time dependent and can be written as follows:

\begin{equation}
\hat H = \mu _1 (t)\hat{a}^ +  \hat{a} + \mu _2 (t) \hat{b}^ +  \hat{b} + g  \psi ^ +  \hat{a}\hat{b} + g^*\psi \hat{a}^ +  \hat{b}^ +  
\label{ham1}
\end{equation}

\noindent
where the molecular field energy is set to zero.
This Hamiltonian different from the one considered in \cite{first} where $A=B$. Here $\hat{a}$ and $\hat{b}$ are anihilation operators of distinct
atomic modes so $[\hat{a}^+,\hat{b}]=0$. Only atomic mode with zero momentum can be described by the process $A_2 \rightarrow A+A$. In the case of
Hamiltonian (\ref{ham1}) the atomic modes $A$ and $B$ become correlated.
 We assume that there is a macroscopic number of molecules that 
does not change considarably during the process so that we can substitute
 $<\hat{\psi}>$ instead of $\hat{\psi}$. Substituting  $g<\psi>^*=\gamma$ into (\ref{ham1}) we get the following atomic Hamiltonian:

\begin{equation}
\hat H = \mu _1 (t)\hat{a}^ +  \hat{a} + \mu _2 (t)\hat{b}^ +  \hat{b} + \gamma \hat{a} \hat{b} + \gamma^*  \hat{a}^ +  \hat{b}^ +  
\label{ham2}
\end{equation}

 Schrodinger's equation leads to the following operator equations:

$$
 i\dot {\hat{a}} = \mu _1 (t)\hat{a} + \gamma^* \hat{b}^ +   
$$ 
\begin{equation} 
 i\dot {\hat{b}^ +}   =  - \mu _2 (t)\hat{b}^ +   - \gamma \hat{a} 
\label{eq1}
\end{equation}
and
\begin{equation}
\begin{array}{l}
 i\dot {\hat{a}^ +}   =  - \mu _1 (t)\hat{a}^ +   - \gamma  \hat{b} \\ 
 i\dot {\hat{b}} = \mu _2 (t)\hat{b} + \gamma^* \hat{a}^+ \\ 
 \end{array}
\label{eq11}
\end{equation}
\noindent
If we make the following change of variable:
\begin{equation}
\begin{array}{l}
 \hat a \to \hat ae^{ - i\int_{}^t {(\mu _1 (t)/2 - \mu _2 (t)/2)dt} }  \\ 
 \hat b \to \hat be^{i\int_{}^t {(\mu _1 (t)/2 - \mu _2 (t)/2)dt} }  \\ 
 \mu (t) = \mu _1 (t)/2 + \mu _2 (t)/2 \\ 
 \end{array}
\label{change}
\end{equation}
\noindent
then  equations (\ref{eq1}),(\ref{eq2}) read:
\begin{equation}
\begin{array}{l}
 i\dot{ \hat{a}} = \mu (t)\hat{a} + \gamma^* \hat{b}^ +   \\ 
 i\dot {\hat{b}^ + }  =  - \mu (t)\hat{b}^ +   - \gamma  \hat{a} \\ 
 \end{array}
\label{eq2}
\end{equation}
\noindent
and
\begin{equation}
\begin{array}{l}
 i\dot {\hat{a}^ +}   =  - \mu (t) \hat{a}^ +   - \gamma  \hat{b} \\ 
 i\dot {\hat{b}} = \mu (t)\hat{b} + \gamma^* \hat{a}^+ \\ 
 \end{array}
\label{eq22}
\end{equation}

In the adiabatic approximation there are no transitions except for the case when adiabaticity is violated near curve crossing points $\mu(t)=0$.
 According to Landau-Zener theory, to estimate transition amplitude we can approximate $\mu(t)$ linearly near the crossing points,
  hence the equation that should be considered are:

\begin{equation}
\begin{array}{l}
 i\dot {\hat{a}} = \beta t \hat{a} + \gamma^* \hat{b}^ +   \\ 
 i\dot {\hat{b}^ +}   =  - \beta t \hat{b}^ +   - \gamma  \hat{a} \\ 
 \end{array}
\label{eq3}
\end{equation}

The corresponding second order differential equation for operator $\hat{a}$ reads:

\begin{equation}
\ddot {\hat a} + (\beta ^2 t^2  - |\gamma|^2  + i\beta ^2 )\hat{a} = 0
\label{eq4}
\end{equation}
\noindent
which is exactly the same operator equation that appears in the case of one mode condensate. The details of its solution can be
 found in  \cite{first}. The difference from the one mode process is that the operator $\hat{a}$ in (\ref{eq3})
  is coupled to operator $\hat{b^+}$ rather than $\hat{a}^+$. 
 The solution in the two mode case can be written in the form:

\begin{equation}
\hat {a}(t) = \phi _c (t)\hat a(t_0 ) + \phi _s (t)\hat b^ +  (t_0 )
\label{sol1}
\end{equation}
\noindent
with initial conditions:

\begin{equation}
\begin{array}{l}
 \phi _c (t_0 ) = 1,\phi _s (t_0 ) = 0 \\ 
 \dot {\phi} _c (t_0 ) = \beta t_0 ,\dot \phi _s (t_0 ) =  - i\gamma^* \\ 
 \end{array}
\label{bc1}
\end{equation}

The average numbers of the atoms in the $A$-mode are:

\begin{equation}
<\hat{a}^+(t)\hat{a}(t)>=n_{st}+n_{sp}
\label{num}
\end{equation}
\noindent
where
\begin{equation}
n_{sp}=|\phi _{s}(t)|^2
\label{nsp}
\end{equation}
\noindent
corresponds to the spontaneous transitions into the $A-$atomic vacuum state. 
This term does not appear if the problem is treated in the mean
 field approximation and is the result of quantum effects \cite{first}, \cite{Vardi8}, \cite{Y14} 

The quantity $n_{st}$ corresponds to the stimulated transitions in the case when atomic states were initially populated:

\begin{eqnarray}
n_{st}	=|\phi _c|^2<\hat{a}^+(t_0)\hat{a}(t_0)>+ |\phi _s|^2 <\hat{b}^+(t_0) \hat{b}(t_0)>+2Re(\phi _s ^* \phi _c <\hat{a}(t_0) \hat{b}(t_0)>)
\label{nst}
\end{eqnarray}

The functions $\phi_c(t)$ and $\phi_s(t)$ are c-functions that are independent solutions of the equation (\ref{eq4}). 
This equation with the initial conditions (\ref{bc1}) appears also in the one mode case which was investigated in \cite{first}, (see appendix).
 Using the results of \cite{first} and substitute it into (\ref{sol1}) we find assimptotics for
 the self-consistent solution of our initial valued problem is as follows:
 
\begin{equation}
\begin{array}{l}
 \phi _s (t \to \infty ) = \frac{{|\gamma|}}{\gamma}\sqrt {\exp (2\pi \lambda )-1} e^{
 \frac{-i3\pi }{4} - iS (|\tau_0|)-iS(\tau) - i\arg \Gamma (i\lambda )}, \\  
 \phi _c (t \to \infty ) = e^{\pi \lambda + iS (|\tau_0|)-iS(\tau)}  \\
 \end{array}
\label{ff1}
\end{equation}
where $\tau =\sqrt{2\beta}t$, $S(\tau)=\tau ^2 /4 -\lambda \ln \tau$ and $\lambda=|\gamma|^2/(2\beta)$
Using this one can derive
\begin{equation}
\begin{array}{l}
 |\phi _s (t \to \infty )|^2  = e^{2\pi \lambda }  - 1 \\ 
 |\phi _c (t \to \infty )|^2 =n_{sp}  = e^{2\pi \lambda }  \\ 
 \end{array}
\label{ff2}
\end{equation}


\noindent
so if initially we have coherent atomic states $|\alpha>|\beta>$ where $\hat{a} |\alpha>=\alpha |\alpha>$ and 
 $\hat{b} |\beta>=\beta |\beta>$ then
\begin{eqnarray}
n_{st}  &=& |\alpha |^2 e^{2\pi \lambda } +|\beta|^2(e^{2\pi \lambda }  - 1) 
 + 2\sqrt {e^{2\pi \lambda }  - 1} e^{\pi \lambda }|\alpha||\beta|\times\nonumber\\
	& & \cos (\frac{3\pi }{4}+2\ S(|\tau _0|)+\arg (\Gamma(i\lambda)) + \arg (\gamma) +\ arg (\alpha ) + \arg (\beta))
\label{nst2}
\end{eqnarray}
\noindent
so, depending on the initial arguments of $\alpha$ and $\beta$ number of produced particles due to the stimulated transitions can be in the range
$ n_{-} < n_{st} < n_{+} $ where

\begin{equation}
n_{\pm}=(|\alpha| e^{\pi \lambda} \pm |\beta|\sqrt{e^{2\pi \lambda}-1})^2
\label{npm}
\end{equation}

Now, let us proceed to the problem of correlations between the two modes. We will investigate the case when initially
 there were no particles in atomic condensates. For this purpose we consider the quadrature phase amplitudes:

\begin{equation}
\hat{X}_{\theta}(t)=(\hat{a}(t)+\hat{b}(t))e^{i\theta} + (\hat{a}^+(t) + \hat{b}^+(t))e^{-i\theta}
\label{X1}
\end{equation}
\noindent
If initially the atomic condensates are in vacuum states then:

\begin{equation}
<\hat{X_{\theta}}^2 (t)>=|\phi _c (t) e^{i\theta}+\phi _s^* (t)e^{-i\theta}|^2
\label{X2}
\end{equation}
and long time after curve crossing event we obtain:

\begin{equation}
<\hat{X_{\theta}}^2 (t \rightarrow \infty)>=|\sqrt{e^{2\pi \lambda}-1}+e^{\pi \lambda -\frac{\imath3\pi}{4}-\imath\arg {\gamma} -2\imath S(\tau)-
 \imath \arg \Gamma(i \lambda)+2\imath \theta}|^2
\label{X23}
\end{equation}

Choosing two orthogonal phase angles $\theta_{+} = 1/2(\frac{3\pi}{4}+\arg {\gamma} +2S(\tau)+\arg \Gamma(i \lambda))$ and
 $\theta_{-} = 1/2(\frac{-\pi}{4}+\arg {\gamma} +2S(\tau)+\arg \Gamma(i \lambda))$ we find:
 
\begin{equation}
 <X_{{\theta}_{\pm}}^2>=|\sqrt{e^{2\pi \lambda}-1} \pm e^{\pi \lambda}|^2
\label{X3}
\end{equation}
\noindent
which is what one gets in the one mode case \cite{first} but with different interpretation. The result (\ref{X3}) means that
 dissociation forms atoms in a two modes squeezed state. On the applications of two mode squeezed states in atomic condensate see for example
\cite{J10}, \cite{Uffe9}, \cite{ss11}.

Finally we would like to discuss another application of our results.
The same Hamiltonian (\ref{ham2}) would appear in three components condensate (that 
differ for example by spin projection) if the following reaction during atomic collisions is possible:

\begin{equation}
C+C \longrightarrow A+B
\label{chC}
\end{equation}
This process leads to a $g\hat{\psi} \hat{ \psi} \hat{a}^+ \hat{b}^+$ term in the Hamiltonian which in the case of large number of $C$ 
atoms can be substituted by 
$g<\hat{\psi} \hat{\psi}>\hat{a}^+\hat{b}^+$.

The time dependent chemical potential crossing can be achived by
 changing adiabatically the shape of the trap. This would produce the change of condensate 
concentration and hence of the chemichal potentials.

If components $A$ and $B$ are distinct by some internal degree of freedom that is not conserved during the reaction
 then the Hamiltonian (\ref{ham1}) should be generalized so that reaction go in all possible channels:

\begin{equation}
\hat{H} = \mu _1 (t) = \hat{a}^+ \hat{a} + \mu _2(t) \hat{b}^+ \hat{b} + \gamma \hat{a} \hat{a} + \gamma ^* \hat{a}^+ \hat{b}^+ + \gamma_a \hat{a}\hat{a}  
  + \gamma _a ^* a^+ a^+ +\gamma _b bb +\gamma_b ^* b^+ b^+
\label{ham4}
\end{equation} 
 
 In this case even if we treat the field of initial condensate as a $c$-number, the resulting Landau-Zener problem in not analytically solvable, since
  all four operator equations
 become coupled. However, if additional symmetry exists, exact asymptotics can be found. For example, if the chemical potentials are equal
  due to some symmetry:
 
\begin{equation}
\mu_1 (t) =\mu _2 (t)
\label{ch2}
\end{equation}
\noindent
  then the equivalent Landau-Zener problem can be solved. For simplicity let us also assume that $\gamma _b = \gamma _a =\gamma_1$. 
  Then the operators evolution is govenered by the following equations:
 
\begin{equation}
\begin{array}{l}
 i\dot {\hat{a}} = t \hat{a}+ \gamma b^+ -2 \gamma^* _1 \hat{a}^+  \\
 i\dot {\hat{b}} = t \hat{b} + \gamma ^* a^+ -2 \gamma^* _1 \hat{b}^+ \\
\end {array}
\label{eq6}
\end{equation}
\noindent
 this is accomplished by the hermitian conjugate equations.
 
If we add the equations in (\ref{eq6}) and denote $\hat{c}= \hat{a}+ \hat{b}$ we get
 
 \begin{equation}
 i\dot{\hat{c}} = t \hat{c} +(\gamma^* - 2\gamma_1 ^*)\hat{c}^+
 \label{eq71}
 \end{equation}
\noindent
 so that for $\hat{c}$ and $\hat{c}^+$ we have a solvable two state Landau-Zener system. Accordingly, we have similar equation for
 operators $\hat{d}$ and $\hat{d}^+$, where $\hat{d}=\hat{a}-\hat{b}$:
 
 \begin{equation}
  i\dot{\hat{d}} = t \hat{d} -(\gamma^* +2\gamma_1 ^*)\hat{d}^+
 \label{eq71}
 \end{equation}
 At the end we can switch back to operators $\hat{a}$ and $\hat{b}$. Such solution would be very unstable in respect of the terms that break the
  symmetry appear in the Hamiltonian given by (\ref{ch2}) since there are states that have the same energy participating in the curve crossing.
   This instability was investigated in a set of articles \cite{Y1}, \cite{Y2}, \cite{Y3}.
   
In conclusion, we considered a time dependent problem of coherent dissociation of molecular condensate into distinct modes beyond
 the mean field approximation. We showed that it can be
 reduced to the case of two distinct atomic modes and molecular mode. We discussed in particular the case when the molecular condensate has zero momentum because in this case the interaction term with the two resulting modes is equal hence the coupling constant $g$ between the molecular and the atomic field becomes $2g$. This fact is very important since it leads to an enhancement in the number of the atomic condensate atoms that exist in the squeezed state. In the case of two modes condensate the corresponding multistate Landau-Zener problem 
 can be solved exactly since the operators evolution equations decouple.  The produced condensates are entangled in a two modes squeezed 
 state. A more general case of the two mode condensate production during a curve crossing could
 also be exactly solvable if additional symmetry is imposed. However such system
  could be very unstable if terms that break the symmetry participate in the Hamiltonian of the system. 
  
  \section{Appendix}
  The solutions of eq.(9) that satisfy the initial conditions (11) are given by 
\begin{equation}
\phi _c(t)=-i\gamma ^* \frac{\phi _1 ^* (t_0) \phi _1 (t) -\phi _2 ^* (t_0) \phi _2(t)}{W(\phi _1,\phi _2)}
\label{pfic1}
\end{equation}
\begin{equation}
\phi _s(t)=2i\gamma ^* \frac{\phi _2 (t_0) \phi _1 (t) -\phi _1 (t_0) \phi _2(t)}{W(\phi _1,\phi _2)}
\label{pfic1}
\end{equation}

\noindent
where $W(\phi _1,\phi _2)=i(2\beta/\gamma)exp(-\pi \lambda /2)$ and 
 
\begin{equation}
\lambda=|\gamma|^2/(2\beta)
\label{lam}
\end{equation}
\noindent
 is Landau-Zener parameter. The  Functions $\phi _1$ and $\phi _2$ are two standard solutions of parabolic cylinder equation with
  asymptotics at $t_0 \longrightarrow -\infty$:

\begin{equation}
\phi _1 (t _0) \sim \frac{1}{\tau _0} \exp(-\frac{\pi}{4} \lambda +i \frac{\pi}{4} +iS(|\tau _0|))
\label{as1m}
\end{equation}
\begin{equation}
\phi _2 (t_0) \sim \frac{1}{\gamma}\sqrt{2\beta}\exp{(-\frac{\pi}{4}\lambda -i\frac{\pi}{4} -iS(|\tau_0|))},
\label{as2m}
\end{equation}
\noindent
 and at $t \longrightarrow +\infty$:

 \begin{equation}
 \phi _1 (t) \sim \frac{2}{|\gamma|} \sqrt{\beta \sinh{(\pi \lambda)}}\exp{(\frac{\pi}{4}\lambda -i \frac{\pi}{2} -iS(\tau) -i arg \Gamma (i \lambda))}
 \label{as1p}
 \end{equation} 
\begin{equation}
\phi _2 (t) \sim \frac{1}{\gamma}\sqrt{2\beta}\exp{(\frac{3\pi}{4}\lambda -i\frac{\pi}{4} -iS(|\tau_0|))}
\label{as1p}
\end{equation}
  
\noindent {\it Acknowledgments}. This work was supported by NSF under the
grant DMR 0072115 and by DOE under the grant DE-FG03-96ER45598. It is our pleasure to thank V.L. Pokrovsky and L.P. Pitaevskii for the fruitful discussions and encouragement.


\begin{references}

\bibitem{first} V.A. Yurovsky, a. Ben-Reuven, P.S. Julienne cond-mat 0108372

\bibitem{second} A. Vardi, V. A. Yurovsky, J. R. Anglin, cond-mat/0105439

 
\bibitem{rev} Franco Dalfovo, Stefano Giorgini, Lev P. Pitaevskii, Sandro Stringari, Rev. Mod. Phys. V71, Issue 3 pp. 463-512  (1999)

\bibitem{Landau}  L.D.Landau, Physik Z. Sowjetunion 2, 46 (1932)

\bibitem{Zener}  C.Zener, Proc. Roy. Soc. Lond. A 137, 696 (1932)

\bibitem{J10} D. Jaksch, J.I. Cirac P. Zoller cond-matt/0110494

\bibitem{mm} K. Goral, M. Gajda, K. Rza$\dot{z}$ewski Phys.Rev.Lett V86, 1397 (2001), cond-mat/0006192

\bibitem{M12} C. Menotti, J.R. Anglin, J.I. Cirac, P. Zoller, Phys. Rev. A, V63, 023601

\bibitem{SUSY13} D. J. Heinzen, R. H. Wynar cond-mat/0110565

\bibitem{brand} S. Brundobler, V. Elser, J. Phys. A: Math.Gen. 26 (1993) 1211-1227

\bibitem{demkov} Yu. N. Demkov, V. I. Osherov, Zh. Exp. Teor. Fiz. 53 (1967) 1589 (Engl. transl. 1968 Sov. Phys.-JETP 26, 916)



\bibitem{LL}  L.D.Landau, E.M. Lifshitz, Quantum Mechanics Pergamon, Oxford

\bibitem{Vardi8} A. Vardi, J.R. Anglin, Phys. Rev. Lett, V86, No4, 568 (2001)

\bibitem{Y14} V. A. Yurovsky cond-mat/0109463


\bibitem{Uffe9} Uffe V. Polsen, K. Molmer, Phys. Rev. A, V63, 023604

\bibitem{ss11} D. Gottesman, J. Preskill, Phys.Rev. A, V63, 022309


\bibitem{Y1} V.A. Yurovsky, A. Ben-Reuven, Phys. Rev. A. V63, 043404 (2001)

\bibitem{Y2} V. A. Yurovsky, A. Ben-Reuven, P.S. Julienne, Y.B. Band, J. Phys. B 32, 1845, (1999)

\bibitem{Y3} V.A. Yurovsky, A. Ben-Reuven, Phys. Rev. A 60, 4561 (1999)

 

\end {references}
\end{document}